# Meißner–Ochsenfeld effect in semiconductor nanostructures with negative-U shells

N.T. Bagraev , N.A. Dovator, L.E. Klyachkin ✉ , A.M. Malyarenko

Ioffe Institute, St. Petersburg, Russia

✉ leonid.klyachkin@gmail.com

**ABSTRACT**
The Meißner–Ochsenfeld effect is demonstrated for the first time at room temperature. The diamagnetic response of a silicon nanostructure with edge channels covered by chains of negative-U dipole boron centers is studied when put in (removed from) an external magnetic field. Measurements of the diamagnetic response were carried out by recording the values of magnetization and generation currents. There is good agreement between the results of measurements of the generated internal magnetic field obtained using a ferroprobe and recording the EMF induced by the occurrence of generation currents in an external magnetic field, which determines the conditions of the mechanism of the nondissipative transport in the edge channels at room temperature, which is caused by their interactions with single carriers through negative-U dipole boron centers. The interrelation of the magnetization hysteresis and the magnitude of the EMF induced by the occurrence of generation currents indicates the possibilities of the electrical registration of the Meißner–Ochsenfeld effect in nanostructures manufactured within the framework of the Hall geometry.

**Funding.** *The work was financed within the framework of the state assignment of the Federal State Unitary Enterprise Ioffe Institute No. FFUG-2024-0039 of the Ministry of Science and Higher Education of the Russian Federation.*

**Citation:** Bagraev NT, Dovator NA, Klyachkin LE, Malyarenko AM. Meißner-Ochsenfeld effect in semiconductor nanostructures with negative-U shells. *Materials Physics and Mechanics*. 2026;53(1): 8–16.
http://dx.doi.org/10.18149/MPM.5412025_2

# Introduction

For twenty years following the discovery of superconductivity, it was believed that a superconductor is a perfect conductor with zero resistance [1]. Only after the discovery of the Meißner–Ochsenfeld effect, which involved identifying an absolute diamagnetic response when a superconductor was placed in an external magnetic field, was the quantum nature of superconductivity established [2]. Subsequent studies enabled the identification of mechanisms leading to the phase state of matter that gives rise to superconductor properties and outlined pathways for realization of the high-temperature superconductors in various compounds and device structures based on them [3–15]. However, the experimental realization of nondissipative carrier transport at high temperatures remains challenging to this day. The main difficulties are associated with studies to controll the magnitude of the local negative correlation energy (negative-U), which arises from the interplay of the Coulomb repulsion and the electron–vibration interaction (EVI) [16,17]. It is precisely the compensation of the Coulomb repulsion via EVI that leads to nondissipative transport and, accordingly, to the emergence of a superconducting state. Moreover, the magnitude of the negative correlation energy (negative-U) practically determines the value of the critical temperature ($T_C$) corresponding to the transition into the superconducting state. It should be noted that the presence of conditions giving rise to negative-U energy (i.e., possibilities for carrier pairing) is not directly linked to the realization of superconductor properties, because disorder in the system of carrier pairs tends to result in dielectric properties of the object rather than nondissipative transport [18]. Therefore, to ensure coherence in nondissipative transport, it has been proposed to use excess single charge carriers whose transfer via tunneling through formed pair centers not only promotes the emergence of the superconductor properties but also enhances them at high temperatures [19,20]. Nevertheless, these predictions have not yet been experimentally realized due to the insufficient magnitude of negative-U energy in most impurity centers and point structural defects. Furthermore, the metastability of the point defects involved in pairing, arising from the presence of the local phonon mode, also leads to decoherence of nondissipative transport and the consequent destruction of the superconductor properties [5,21]. In this case, the use of hybrid structures based on topological semiconductors and superconductors appears to be promised [22]. In particular, semiconductor nanostructures with the edge channels covered by the chains of the negative-U centers are of great interest, as they promote segmentation into regions containing single charge carriers under the compensation of the electron–electron interaction. This opens new possibilities for nondissipative transport of single charge carriers due to their interaction with the negative-U dipole boron centers. It should be noted that segment of the edge channel containing the single charge carrier can be regarded as the Andreev molecule [23,24], whose characteristics allow the study of macroscopic quantum effects [25].

The conditions described above to provide nondissipative transport via local interaction of single carriers with negative-U dipole boron centers covering the edge channels make them promising for studying quasi-one-dimensional superconductors at high temperatures, up to room temperature. In the present work, the identification of the aforementioned conditions enabling superconductivity at room temperature is demonstrated through the study of the diamagnetic response of the silicon nanostructures in shells consisting of chains of the negative-U dipole boron centers.

## Method

The investigated device structure was the Hall bridge based on a silicon nanostructure (Fig. 1). An ultra-narrow silicon p-type quantum well, which forms the basis of the silicon nanostructure, was created on the surface of the n-type (100) silicon wafer. The Hall bridge was boron-doped from the gas phase with a concentration of $5 \times 10^{21}$ cm$^{-3}$, which is measured by the secondary ion mass spectrometry (SIMS). It was found that within the ultra-shallow (8 nm) boron diffusion profile in silicon, self-ordering effects arise due to the parity of various diffusion mechanisms [26]. As a result, enhanced boron diffusion is observed along certain crystallographic directions, ultimately leading to the formation of chain-like boron-center structures via vacancy-drag effects. In particular, this results in the formation of boron chains oriented along the [011] direction and, accordingly, the alignment of the Hall bridge along this crystallographic axis. These boron chains thus

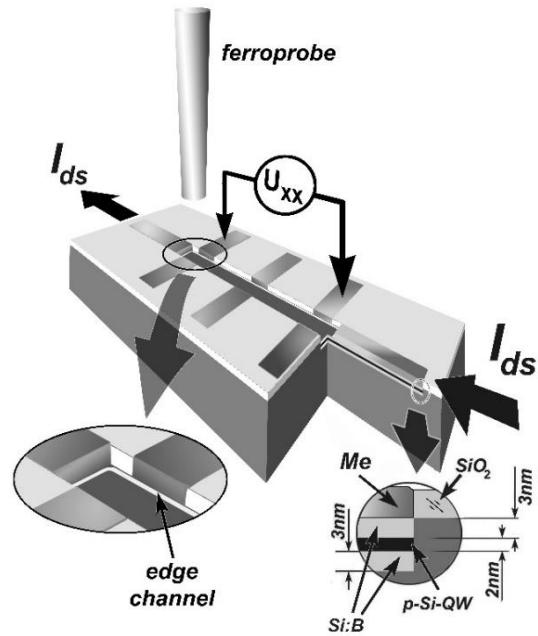

**Fig. 1.** A Hall bridge based on the silicon nanostructure prepared on the surface of (100) silicon wafer, which contain the topological edge channel oriented along the [011] direction with single charge carriers that occupy the separate regions, pixels, because of the compensation of electron-electron interaction. During the experiments, the ferroprobe of the magnetometer was directed perpendicular to the sample surface

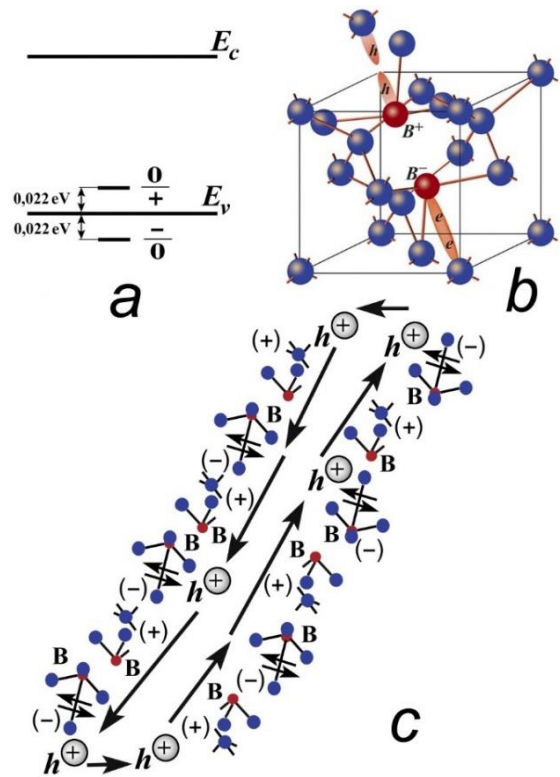

**Fig. 2.** Energy levels (a) of a negative-U dipole boron center (b); model of single charge carrier tunneling transport along a chain of the negative-U dipole centers (c)

confine the obtained quantum well. Given the high concentration, the distance between these chains in the near-surface layer of the nanostructures, including within the barriers that define the quantum well, corresponds to the spacing between boron centers within a single chain – 2 nm (Fig. 1). This important result is largely a consequence of the possible reconstruction of boron centers along the [111] direction upon changes in their charge/spin state (Fig. 2). In turn, the possibility of such reconstruction, as indicated above, is the reason for the formation of negative-U dipole centers due to the emergence of the local phonon mode at the extremely high concentration of boron centers [26]. Thus, the above serves as the foundation for a technology to produce the Hall bridges with a specific crystallographic orientation, which enables the confinement of the edge channels in nanostructures by the chains of the negative-U dipole centers that act as an energy reservoir for nondissipative carrier transport.

It should be noted that chains consisting of paired centers were first proposed by A.Yu. Kitaev [27,28] for exploring possibilities in quantum-computing operations (Fig. 3). However, in the absence of an internal energy reservoir within a spin loop composed of parallel chains of paired centers, the conditions for nondissipative carrier transport are not met. The reason for this is the increasing role of electron–electron interaction. This main obstacle to spin-dependent transport in an interference loop can, as it turns out, be eliminated if negative-U centers are used as the chain components.

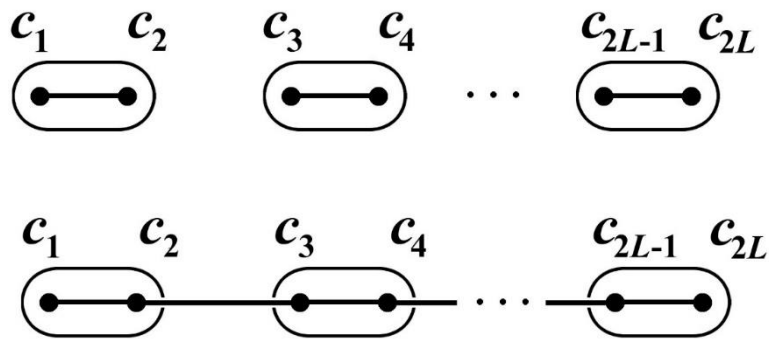

**Fig. 3.** Two possible types of pairing between adjacent dipole centers. Based on [27,28]

If the paired centers within the chains confining the silicon quantum well are negative-U dipole centers, then the energy reservoir for nondissipative carrier transport inside a spin loop composed of parallel chains is based on the unpairing/pairing reaction of a negative-U dipole center [29].

In this case, due to the presence of negative-U energy, singly charged centers decay into doubly charged and empty centers (Fig. 2(a)). Predominantly in tetrahedral systems, this formation of the dipole centers occurs along the [111] axis, provided that the chain is oriented along the [011] axis. Thus, the chains break down into a system of dipole centers, except for the terminal centers, which contain the single charge carrier (hole) near the contacts (Fig. 2(c)). During the passage of a longitudinal current, the single charge carrier tunnels through the dipole centers within the negative-U chain system:

$$D^- + h \Rightarrow D^0$$
$$D^+ \Rightarrow D^0 + h \qquad (1)$$
$$2D^0 \Rightarrow D^+ + D^-$$

As a result of the described reactions, when a carrier tunnels, it moves to the next dipole center, having received energy from the chain. The subsequent decay of the neutral components of the dipole center is accompanied by the reverse release of energy into the chain simultaneous with its reconstruction. Thus, the exchange of energy between the carrier and the chain provides the condition for nondissipative transport during tunneling [29]. Accordingly, the tunneling processes described above are accompanied by distortions within the chain system. The criterion for realizing nondissipative transport is the evaluation of the ratio between the decay time ($\tau_1$) of the neutral dipole states and the tunneling time ($\tau_2$): $\tau_1 = h/\Delta E$ = 6.62 × 10$^{-34}$/44 meV, where 44 meV corresponds to the reduction in entropy during the decay of dipole negative-U centers; $\tau_1$ = 8 x 10$^{-14}$ s; $\tau_2 = h/\Delta E$ = h/($I^2R$) = h/($I^2h/e^2$).

Given the values of the generation current used and the fact that the resistance of a dipole-center cell corresponds to the quantum of resistance, $\tau_1 < \tau_2$, which ensures the conditions for nondissipative transport at $T$ = 300 K. It should also be noted that the two-dimensional carrier density in such edge channels, according to the Hall measurements, is 3 × 10$^{13}$ m$^{-2}$ [30].

Given the realization of nondissipative transport of single holes due to energy exchange with chains of dipole negative-U boron centers, the size of the spin-dependent transport loop is limited to a strip of dimensions 2 nm × 16 μm. It should be noted that the quantum interference regime does not depend on the loop's shape but is determined solely by its area. This area primarily defines the magnetic field value corresponding to optimal quantum interference conditions: $\Delta\Phi = \Delta BS = \Phi_0 = h/e$ = 4 × 10$^{-15}$ Wb, where $S$ is the area of the interference loop; $\Delta B$ is the period of quantum oscillations as a function of the external magnetic field; $\Phi_0$ is the magnetic flux quantum.

It should be noted that this condition is consistent with the Landau quantization conditions, which, accordingly, leads to a stepwise change in the areas of interference loops when the magnetic subband index changes. Taking into account the parameters of the edge channel region occupied by a single hole in the silicon nanostructure, the optimal magnetic field value corresponding to quantum interference is 0.124 T [30].

Thus, the characteristics of the investigated nanostructure provide the conditions for nondissipative carrier transport, which has enabled the observation of macroscopic quantum effects at high temperatures up to room temperature [30], and defines the prospects for using chains of negative-U dipole centers to observe the Meißner–Ochsenfeld effect at $T$ = 300 K and, correspondingly, to identify their superconductor properties.

## Results and Discussion

The Meißner–Ochsenfeld effect was observed in the study of the silicon nanostructure used, both by measuring the diamagnetic response and through measurements of the longitudinal electromotive force (EMF) induced by occurrence of generating currents when the structure was introduced into an external magnetic field.

As noted above, the transport of single charge carriers in the edge channels of nanostructures confined by the chains of the negative-U dipole centers is accompanied by the generation of a magnetic field [30]. Moreover, under an external magnetic field oriented perpendicular to the plane of the nanostructure's edge channel, a magnetic field arises inside the sample that is equal in magnitude to the external field but opposite in direction (Fig. 1) [2]. That is, the measured field response when the structure is introduced into an external magnetic field (Fig. 4) is similar to the data from studies of the Meißner–Ochsenfeld effect in superconductivity and corresponds to the manifestation of absolute diamagnetism. These results are quite understandable if one considers the similarity between the characteristics of chains of negative-U dipole centers and quasi-one-dimensional superconductors. The use of a ferroprobe magnetometer made it possible to detect a reversal in the sign of the magnetic response $\Delta B$, thereby demonstrating that a magnetic field with the opposite sign appears within the sample. It should be noted that the sample was put in an external magnetic field of +15 nT, and accordingly, the diamagnetic response was identified by measuring the magnetic field value within the investigated structure, which was -15 nT.

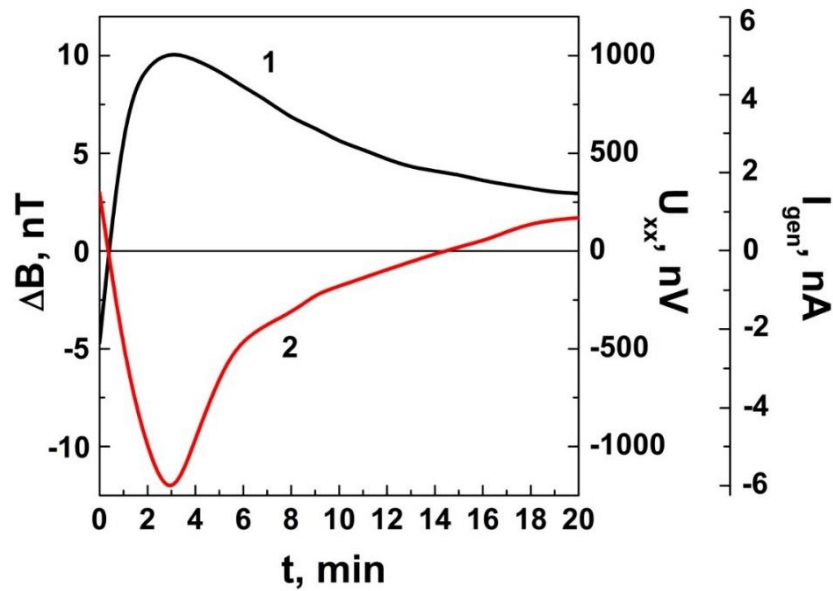

**Fig. 4.** EMF measured between the XX contacts of the silicon nanostructure during insertion (1) and removal (2) of the sample from an external magnetic field of 15 nT. $T$ = 300 K

The experimental setup used was a five-layer magnetically shielded cylindrical chamber. All inner cylinders were made of M-79 permalloy, while the outer cylinder (with a diameter of 50 cm and length of 70 cm) was made of ARMCO steel. The cylinders are closed by lids fabricated from the same material. Inside the innermost cylinder, a Helmholtz coil system powered by a stabilized (adjustable) current source is installed. This system is used both to generate a uniform internal magnetic field and to compensate for the residual magnetic field arising from the penetration (into the shield) of the external laboratory field and the residual magnetization of the permalloy shells. The shielding factors obtained (in the frequency range 0–1 Hz) were 700 for the longitudinal (along the shield axis) and 5000 for the transverse (perpendicular to the shield axis) components of the magnetic field. During the experiments, the magnitude of magnetic induction variations inside the setup was monitored using a cesium-vapor quantum magnetometer and practically did not exceed ± 0.1 nT.

In addition to the above, monitoring the response of the magnetic moment to changes in the external magnetic field enables direct measurement of the generation currents arising within the sample. It is important to note that the EMF generated in the interference loop is directly correlated with the magnitude of the magnetic field inside the sample when it is put in an external magnetic field (Fig. 4). This allows for a significant improvement in the accuracy of the determined magnetic field, and electrical measurements also provide a convenient method for characterizing macroscopic quantum phenomena.

As mentioned above, the magnetic field magnitude inside the sample can be monitored by measuring the EMF between contacts, for example, between the XX contacts of the Hall bridge (Fig. 1). Considering the parameters of the pixels containing single carriers in the edge channel, 16.6 μm × 2 nm, their number between the XX contacts spaced 2 mm apart can be determined. Taking into account that the resistance of a single pixel $R_{pix}$, which represents a quantum box containing a single charge carrier, is $h/e^2$ = 25812 Ω, then with 125 pixels between the XX contacts, the total resistance R for their parallel connection is 200 Ω. This allows the generation current induced by the external magnetic field to be determined (Fig. 4): $I_{gen} = U_{xx}/R$.

Thus, considering that the generation of the internal magnetic field can be described within the framework of the Faraday electromagnetic induction, and taking into account the parameters of the studied structure, particularly the area of the loop between the XX contacts (Fig. 1) where the magnetic flux $S$ = 1 μm × 2 mm [18] is localized, it can be concluded that there is good agreement between the electrical and magnetic measurements of the generated internal magnetic field.

A control experiment to study the correlation between the magnetic and electrical measurements of the external magnetic field described above involves measuring the EMF response between the XX contacts when the sample is removed from the external magnetic field (Fig. 4). The long-duration transient processes accompanying the generation of the internal magnetic field upon insertion/removal of the sample from the external magnetic field indicate the metastability of the Meißner–Ochsenfeld effect, resulting from the recharging within the system of parallel-connected quantum resistances and capacitances of pixels in the edge channels.

## Conclusion

The Meißner–Ochsenfeld effect has been observed for the first time at room temperature. The diamagnetic response of a silicon nanostructure with the edge channels confined by the chains of the negative-U dipole boron centers has been studied upon insertion into (and removal from) an external magnetic field. It has been demonstrated that there is good agreement between the results of measurements of the generated internal magnetic field obtained using a ferroprobe and the EMF induced by the onset of generating currents in the external magnetic field. The observation of magnetization hysteresis due to the Meißner–Ochsenfeld effect at room temperature indicates the presence of superconductor properties induced by the interaction of the single charge carriers with the chains of the negative-U dipole centers in the shells of the nanostructure's edge channels.

## CRediT authorship contribution statement

**Nikolai T. Bagraev**: writing – review & editing, conceptualization; **Nikolai A. Dovator**: investigation; **Leonid E. Klyachkin**: writing – original draft, investigation, data curation; **Anna M. Malyarenko**: investigation, supervision

## Conflict of interest

The authors declare that they have no conflict of interest.